\documentclass[twocolumn,showpacs,preprintnumbers,amsmath,amssymb]{revtex4}
\usepackage{amsmath,amssymb,graphics,epsfig,subfigure}
\usepackage{color}
\usepackage{hyperref}

\newcommand{\tcb}{\textcolor{blue}}

\newcommand{\sign}{\text{sign}}

\begin{document}

\thispagestyle{empty}

\begin{center}

\title{The universal topological charge of black hole photon spheres in higher dimensions}

\date{\today}
\author{Jun-Lei Chen, Shan-Ping Wu, Shao-Wen Wei \footnote{E-mail: weishw@lzu.edu.cn, corresponding author}}

\affiliation{$^{1}$ Key Laboratory of Quantum Theory and Applications of MoE, Gansu Provincial Research Center for Basic Disciplines of Quantum Physics, Lanzhou University, Lanzhou 730000, China\\
	$^{2}$Lanzhou Center for Theoretical Physics, Key Laboratory of Theoretical Physics of Gansu Province, School of Physical Science and Technology, Lanzhou University, Lanzhou 730000, People's Republic of China,\\
	$^{3}$Institute of Theoretical Physics $\&$ Research Center of Gravitation,
	Lanzhou University, Lanzhou 730000, People's Republic of China}

\begin{abstract}
A recently developed topological approach offers novel insights into photon spheres, which are fundamental to the formation of black hole shadows. In this study, we extend this topological analysis to higher-dimensional, static, spherically symmetric, and asymptotically flat black holes. By examining the asymptotic properties of the vector field associated with the photon spheres, we demonstrate that their topological charge is consistently -1. This result is a dimensionally independent invariant, guaranteeing the existence of at least one standard (unstable) photon sphere outside the event horizon. We further explore this conclusion by analyzing two distinct regular black hole solutions derived from pure gravity theory, confirming that the topological charge remains -1 irrespective of the spacetime dimension. These results provide a robust and universal characterization of photon spheres in higher-dimensional spacetimes.
\end{abstract}


\maketitle
\end{center}

\section{Introduction}
\label{secIntroduction}

Black holes are one of the most profound predictions of general relativity. For much of their history, they remained objects of purely theoretical fascination, largely due to their unique causal structure, which precludes direct observation. The presence of event horizons, alongside formidable theoretical challenges such as the information-loss paradox \cite{Hawking} and the problem of spacetime singularities \cite{Penrose}, historically confined black holes to the realm of theory. This paradigm was revolutionized by a series of landmark observational achievements, including the detection of gravitational waves from binary black hole mergers by the LIGO-Virgo collaboration \cite{Abbott1} and the imaging of supermassive black holes in M87 and the Galactic Center by the Event Horizon Telescope (EHT) \cite{Akiyama1,Akiyama3}. These discoveries provided definitive confirmation of their existence. Central to the indirect observation of black holes are photon spheres (PS), unstable orbits for light that are fundamental to key observable signatures. PSs critically shape the appearance of black hole shadows \cite{Chandrasekhar} and govern the characteristic frequencies of quasi-normal modes \cite{Cardoso}.

String theory predicts the existence of extra dimensions, with five-dimensional extremal black holes playing a key role in understanding black hole entropy \cite{Strominger}. In parallel, Anti-de Sitter/conformal field theory (AdS/CFT) establishes a connection between $D$-dimensional black holes and $(D-1)$-dimensional quantum field theory \cite{Maldacena}. This framework enables the study of complex quantum field theory problems by leveraging the properties of high-dimensional black holes, thereby stimulating significant interest in their study. PSs are closely linked to shadow structure, facilitating the investigation of high-dimensional black hole characteristics. For instance, research on various black hole solutions and gravitational theories \cite{Amarilla,Papnoi,Singh,Amir,Eiroa,Belhaj,TsukamotoKitamura,NovoCunha} has demonstrated that extra dimensions influence the PS, which in turn reduces the shadow size. By integrating these findings with astronomical observations, constraints can be placed on theoretical parameters. Consequently, investigating the PS of high-dimensional black holes is both meaningful and necessary.

Recently, the topological approach has emerged as a novel perspective in the study of PSs, attracting significant attention. This approach began with Cunha, Berti, and Herdeiro \cite{Cunha1}. Based on the Brouwer degree of a continuous map, they demonstrated that the light rings (LRs) of ultra-compact objects appear in pairs. Further extending this work, Cunha and Herdeiro \cite{Cunha2} proved that, for each sense of rotation, at least one standard LR exists outside the horizon of a four-dimensional stationary, axisymmetric, asymptotically flat black hole. They achieved this by calculating the winding number of the vector field defined by the effective potential in the orthogonal $(r, \theta)$ space, a generalization of their earlier work on the LRs of ultra-compact objects.

Inspired by this, and employing with Duan's $\phi$-mapping topological current theory \cite{Duan1,Duan2}, Wei proposed a similar topological method \cite{Wei1} for studying the PSs of general four-dimensional static spherically symmetric black holes under three asymptotic conditions: asymptotically flat, AdS, and dS. The results showed that the total topological charge is always equal to -1 in asymptotically flat, AdS, and dS spacetimes, indicating the presence of at least one standard PS outside the black hole's horizon. Moreover, the evolution from a Dyonic black hole to a naked singularity shifts the total topological charge from -1 to 0 \cite{Wei1}. As the horizon disappears, the two PSs, initially separated by the horizon, converge and annihilate.

The topological approach offers the advantage of bypassing specific field equations while providing general conclusions about the topology of PSs or LRs. It also facilitates the intuitive distinction between black holes and horizonless ultra-compact objects, as they belong to different topological classes. Study on the topology of PSs or LRs has been extensively applied to different black hole backgrounds including extreme black holes, Einstein-Maxwell-Dilaton black holes \cite{GuoGao,Wu,Junior1,Junior2,Hosseinifar1,Liu1,Afshar1,Liu2,Cunha3,Moreira,Xavier,Afshar2,Shahzad2,Hosseinifars}. The approach has proven useful in further studies, including those on timelike circular orbits \cite{Wei2}, critical points \cite{Wei3}, and black hole solutions \cite{Wei4}. Additionally, studies of Refs. \cite{Wei1,Junior2} indicated that the topology of PSs or LRs is closely tied to the asymptotic structure outside the black hole's horizon. In Ref. \cite{Junior2}, the effect of a dilaton coupling, which modifies the asymptotic behavior of spacetime, was investigated. It was found that dilaton coupling induces a topological transition in the topological charge. Similarly, since the spacetime dimension also influences the asymptotic behavior of the spacetime, we expect to investigate whether the spacetime dimension affects the topological charge. Our study explores the topology of PSs in high-dimensional asymptotically flat black holes, focusing on the case of static spherical symmetry for simplicity.

The paper is organized as follows. In Sec. \ref{tc}, we introduce the topological current and charge based on Duan's $\phi$-mapping topological current theory. In Sec. \ref{tchd}, we provide a brief overview of the PSs of high-dimensional static spherically symmetric (SSS) black holes and present a theoretical derivation of the topological number of PSs for high-dimensional SSS asymptotically flat black holes. In Sec. \ref{tchdrbh}, we present two regular black hole solutions constructed from pure gravity to test our conclusion. Finally, in Sec. \ref{Conclusion}, we summarize and discuss our results.

\section{Topological approach}
\label{tc}

In order to study the topology of PSs, we first introduce the corresponding topological current and topological charge based on Duan's $\phi$-mapping topological current theory.
		
According to Duan's $\phi$-mapping topological current theory, we can construct the topological current
\begin{equation}
    j^{\mu }=\frac{1}{2\pi}\epsilon^{\mu \nu \rho}\epsilon_{ab}\partial_\nu n^{a}\partial_\rho n^{b},\qquad\mu,\enspace\nu,\enspace\rho=0,\enspace 1,\enspace 2.
\end{equation}
Here $\partial_\nu =\partial / \partial x^{\nu}, x^\nu=(t,\enspace r,\enspace\theta )$. $n^{a}$ is the unit vector,
\begin{equation}
    n^{a}=\frac{\phi^{a}}{||\phi||},\qquad a=1,\enspace 2,
\end{equation}
with $\phi^{1}=\phi^{r}$, $\phi^{2}=\phi^{\theta}$ and $||\phi||=\sqrt{(\phi^{1})^{2}+(\phi^{2})^{2}}$. It is easy to check the conservation of the topological current
\begin{equation}
    \partial_\mu j^{\mu}=0,
\end{equation}
where $j^{0}$ is the charge density. Then we can obtain the topological charge $Q$ at given region $\Sigma$
\begin{equation}
    Q=\int_{\Sigma}  j^{0}\,d^{2}x.
\end{equation}
Next, we explore the inner structure of topological charge. Using the Jacobi tensor and the two-dimensional Laplacian Green function, we get
\begin{equation}
    j^{\mu }=\delta ^{2}(\phi)J^{\mu}\left(\frac{\phi}{x}\right),
\end{equation}
where $\epsilon^{ab}J^{\mu}(\phi/x)=\epsilon^{\mu\nu\rho}\partial_{\nu}\phi^{a}\partial_{\rho}\phi^{b}$. Obviously, $j^{\mu}$ is not equal to 0 only at zero points of $\phi ^{a}$. Let the $m$-th solution of $\phi ^{a}(\vec{x})=0$ be $\vec{x}=\vec{z}_{m}$. By the nature of $\delta $-function, one can obtain
\begin{equation}
    j^{0}=\delta^{2}(\phi)J^{0}\left(\frac{\phi}{x}\right)
         =\sum_{m = 1}^{N} \beta _{m} \eta _{m} \delta ^{2}(\vec{x}-\vec{z}_{m}),
\end{equation}
where the positive Hopf index $\beta_{m}$ is the number of loops made in vector space of $\phi$ when $x^{\mu}$ makes one loop around the zero point $z_{m}$ and the Brouwer degree $\eta_{m}=\sign[J^{0}(\phi/x)|_{\vec{x}=\vec{z}_{m}}]= \pm 1$. So we can obtain the topological charge $Q$ at given region $\Sigma$
\begin{equation}
    Q=\int_{\Sigma}  j^{0}\,d^{2}x
     =\sum_{m=1}^{N} \beta_{m} \eta_{m}
     =\sum_{m=1}^{N} w_{m}.
\end{equation}
Here, $w_{m}$ represents the winding number of the zero point $z_{m}$ contained in $\Sigma$. $Q$ is defined as the sum of the winding numbers of all zero points of $\phi$ within $\Sigma$ for a given parameter region. $w_{m}$ reflects local topological properties, whereas $Q$ encapsulates the global topological properties.

For convenience, we provide numerical calculation method for winding number and topological charge. The value of $w_{m}$ is independent of the shape of the integration path, requiring only that the path encloses the zero point $z_{m}$. Similarly, the value of $Q$ is independent of the path's shape, as long as the path encloses all the zeros within the given parameter region $\Sigma $. It is advantageous to express the parametric equation of the integration path as follows:
\begin{equation}
    \left\{
        \begin{aligned}
            r&=a\cos\vartheta+r_{0},\\
            \theta&=b\sin\vartheta+\frac{\pi}{2}.
        \end{aligned}
    \right. \label{rthetavartheta}
\end{equation}

The value of $w_{m}$ and $Q$ can be calculated by the following equation
\begin{equation}
    \frac{\Omega (2\pi)}{2\pi}=\frac{1}{2\pi}\int_{0}^{2\pi} \epsilon_{ab} n^{a}\partial_{\vartheta}n^{b} \,d\vartheta ,\label{omegapi}
\end{equation}
where $\Omega$ is the deflection angle of $\phi$ in the ($r$, $\theta$) plane and
\begin{equation}
    \Omega(\vartheta^{\prime})=\int_{0}^{\vartheta^{\prime}} \epsilon_{ab} n^{a}\partial_{\vartheta}n^{b} \,d\vartheta.\label{omegavartheta}
\end{equation}

\section{Topological charge in high dimensions}
\label{tchd}

We consider $D$-dimensional SSS black holes, which have the following line element
\begin{equation}
ds^{2}=-f(r)dt^{2}+\frac{dr^{2}}{g(r)}+h(r)d\varOmega^{2}_{D-2}, \label{metric}
\end{equation}
where $D\ge 5$. The radius $r_{\text{h}}$ of the black hole event horizon is the largest root of $f(r)=0$ or $g(r)=0$.

Now let us turn to the PSs. Note that
\begin{equation}
\begin{aligned}
d\varOmega^{2}_{D-2}=&d\theta^{2}_{1}+\sin^{2}\theta_{1}d\theta^{2}_{2}+\sin^{2}\theta_{1}\sin^{2}\theta_{2}d\theta^{2}_{3}\\&+\cdots+\sin^{2}\theta_{1} \cdots \sin^{2}\theta_{D-3}d\theta^{2}_{D-2},
\end{aligned}
\end{equation}
where $\theta_{k}\in[0,\ \pi]$  ($k=1,2,\cdots,D-3$) and $\theta_{D-2} \in [0,\ 2\pi)$. Due to the spherical symmetry, we consider the equatorial hyperplane defined by
\begin{equation}
\theta_{1}=\theta_{2}=\cdots=\theta_{D-3}=\frac{\pi}{2}.
\end{equation}
Then, the line element Eq. (\ref{metric}) reduces to
\begin{equation}
ds^{2}=-f(r)dt^{2}+\frac{dr^{2}}{g(r)}+h(r)d\theta^{2}_{D-2}.\label{reducel}
\end{equation}
Despite the fact that we take values for $\{\theta_{n}\}$, the information of dimension $D$ is retained in $f(r)$ and $g(r)$.

Two conserved quantities can be written as
\begin{equation}
        \begin{aligned}
            &L=-\frac{\partial(-\frac{1}{2}g_{ab} \dot x^{a} \dot x^{b})}{\partial\dot \theta_{D-2}}=h(r)\dot \theta_{D-2},\\
            &E=\frac{\partial(-\frac{1}{2}g_{ab} \dot x^{a} \dot x^{b})}{\partial\dot t}=f(r)\dot t,
        \end{aligned}
\end{equation}
where the dot indicates the derivative with respect to the affine parameter $\lambda$. $E$ and $L$ are the energy and angular momentum of photon, respectively corresponding to the Killing vector fields $\partial_t$ and $\partial_{\theta_{D-2}}$.

Considering $-\frac{1}{2}g_{ab} \dot x^{a} \dot x^{b}=0$, the radial motion on the equatorial hyperplane plane can be obtained:
\begin{equation}
\dot r^{2} +V_{\text{eff}}=0,
\end{equation}
where the effect potential $V_{\text{eff}}=g(r)(L^{2}/h(r)-E^{2}/f(r))$.

With the following conditions, we can solve for the position of the PSs
\begin{equation}
V_{\text{eff}}=0,\quad \partial_{r} V_{\text{eff}}=0.\label{veff}
\end{equation}
For the reduced line element (\ref{reducel}), we obtain by solving Eqs. (\ref{veff})
\begin{equation}
h(r)f^{\prime}(r)-f(r)h^{\prime}(r)=0,\label{rph}
\end{equation}
where the prime indicates the derivative with respect to $r$. The solutions of the above equation correspond to the position of the PSs.

To study the topology of the PSs of high-dimensional SSS asymptotically flat black holes, it is essential to begin with their metric functions. Previous studies \cite{Wei1,Junior2} have shown that the topology of the PSs is directly related to the asymptotic behavior of the black hole's metric functions. Therefore, the topology of the PSs of high-dimensional SSS asymptotically flat black holes can be analyzed by examining the asymptotic behavior of the metric functions. Specifically, the asymptotic behaviors of these functions at infinity are expressed as follows:
\begin{eqnarray}
    f(r) &\sim& 1-\frac{m}{r^{D-3}}+\mathcal{O}\left(\frac{1}{r^{D-2}}\right),\label{fff}\\
    g(r) &\sim& 1-\frac{m}{r^{D-3}}+\mathcal{O}\left(\frac{1}{r^{D-2}}\right),\\
    h(r) &\sim& r^{2}.\label{hhh}
\end{eqnarray}
Analogous to the four-dimensional case, we employ the vector field $\phi = (\phi^r, \phi^\theta)$
\begin{equation}
\phi^{r}=\frac{\partial_{r}H}{\sqrt{g_{rr}}}=\frac{r f^{\prime}-2f}{2r^{2}\sin\theta},\; \phi^{\theta}=\frac{\partial_{\theta}H}{\sqrt{g_{\theta \theta}}}=-\frac{\sqrt{f}\cos\theta}{r^{2}\sin^{2}\theta},\label{phirtheta}
\end{equation}
where $H=\sqrt{-g_{tt}/g_{\theta_{D-2}\theta_{D-2}}}$ and $\theta \equiv \theta_{1}$. To study the topology of the PSs, we retain the parameter $\theta$ here. The zero points are clearly located at $\phi=(0,0)$. The parameter $\theta$ causes the zero points to lie on $\theta = \pi/2$, facilitating the identification of these points. Each zero point corresponds to a PS. If we consider a zero point as a topological defect, we can apply Duan's $\phi$-mapping topological current theory to determine the topological charge.

\begin{figure}[htbp]
    \center{
    \includegraphics[width=7.5cm]{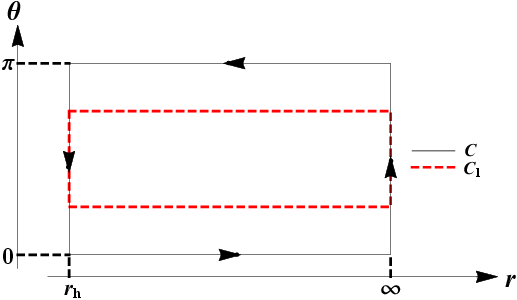}}
  \caption{Two Types of limit boundaries in the ($r$, $\theta$) plane. The black arrow indicates that counterclockwise is the positive direction. $C$ is the union of  black solid line segments: $\{l_{1}:r\to +\infty, 0 \le \theta \le \pi \} \cup \{l_{2}:\theta\to\pi^{-}, r_{\text{h}} < r < +\infty \} \cup \{l_{3}:r\to r^{+}_{\text{h}}, 0 \le \theta \le \pi \} \cup \{l_{4}:\theta\to 0^{+}, r_{\text{h}} < r < +\infty \}$. $C_{1}$ is the union of red dashed line segments: $\{l^{1}_{1}:r\to +\infty, \theta_{0} \le \theta \le \pi-\theta_{0} \} \cup \{l^{1}_{2}:\theta = \pi-\theta_{0}, r_{\text{h}} < r < +\infty \} \cup \{l^{1}_{3}:r\to r^{+}_{\text{h}}, \theta_{0} \le \theta \le \pi-\theta_{0} \} \cup \{l^{1}_{4}:\theta=\theta_{0}, r_{\text{h}} < r < +\infty \}$, with $\theta_{0} \in (0, \pi/2)$.}
  \label{Fig.1}
\end{figure}

To ensure that all zeros are accounted for in calculating the total topological charge, we examine four limiting cases: $r \to r_{\text{h}}^+$, $\theta \to 0^+$, $\theta \to \pi^-$, and $r \to +\infty$, corresponding to $l_{3}$, $l_{4}$, $l_{2}$, $l_{1}$, respectively. $\phi^{r}_{l_{i}}$ and $\phi^{\theta}_{l_{i}}$ represent the components of $\phi$ along line segment $l_{i}$, while $\Omega_{l_{i}}$ represents the deflection angle of $\phi$ along line segment $l_{i}$. To facilitate intuitive understanding, we describe two contours $C$ and $C_1$ containing limit boundaries in ($r$, $\theta$) plane in Fig. \ref{Fig.1}. The contour $C$ is the union of four segments $l_1$--$l_4$, defined as $l_1: r\to+\infty,\ 0\le\theta\le\pi$; $l_2: \theta\to\pi^-,\ r_{\text{h}}<r<+\infty$; $l_3: r\to r_{\text{h}}^+,\ 0\le\theta\le\pi$; and $l_4: \theta\to0^+,\ r_{\text{h}}<r<+\infty$. The contour $C_1$ is similarly composed of $l^1_1$--$l^1_4$ with $l^1_1: r\to+\infty,\ \theta_0\le\theta\le\pi-\theta_0$; $l^1_2: \theta=\pi-\theta_0,\ r_{\text{h}}<r<+\infty$; $l^1_3: r\to r_{\text{h}}^+,\ \theta_0\le\theta\le\pi-\theta_0$; and $l^1_4: \theta=\theta_0,\ r_{\text{h}}<r<+\infty$, where $\theta_0\in(0,\pi/2)$. By counting the change of the direction of $\phi$ along this curve in the $(r, \theta)$ plane, we can determine the total topological charge. Here we take the counterclockwise direction as positive, and this convention will be followed throughout the discussion.

Next, we aim to examine the behaviors of the vector at these boundaries by making use of (\ref{fff})-(\ref{hhh}).

When $r\xrightarrow{}r^{+}_{\text{h}}$, we have
\begin{equation}
    \begin{aligned}
        \phi^{r}_{l_{3}}(r\xrightarrow{}r^{+}_{\text{h}}) &= \frac{ f^{\prime}|_{r=r_{\text{h}}}}{2{r_{\text{h}}}\sin\theta} \ge 0\ ,\\
        \phi^{\theta}_{l_{3}}(r\xrightarrow{}r^{+}_{\text{h}}) &\rightarrow 0,   \end{aligned}
\end{equation}
where $f^{\prime}|_{r=r_{\text{h}}}>0$ and $f(r_{\text{h}})=0$ are used. The variation in $\theta$ only influences the magnitude of $\phi^{r}_{l_{3}}$, without affecting its sign. $\phi_{l_{3}}^{\theta}$ remains horizontal to the right on the $(r,\theta)$ plane, i.e., $\Omega_{l_{3}}=0$, and thus $\Delta\Omega_{l_{3}}=0$. $\Delta\Omega_{l_{i}}$ represents the change in $\Omega_{l_{i}}$ along the positive direction.

Considering $\theta\xrightarrow{}0^{+}$ and $\theta\xrightarrow{}\pi^{-}$, we have
\begin{equation}
    \begin{aligned}
        (r f^{\prime}-2f)|_{r=r_{\text{h}}} = & r_{\text{h}}f^{\prime}|_{r=r_{\text{h}}},\\
        (r f^{\prime}-2f)|_{r\xrightarrow{}+\infty}=&-2. \label{rff}
    \end{aligned}
\end{equation}
The finiteness of $f$ and $f^{\prime}$ at a finite location outside the horizon and the asymptotic behaviors in Eq. (\ref{rff}) ensure that $r f^{\prime}-2f$ does not diverge outside the horizon. By selecting a suitable convergence, we get
\begin{equation}
    \begin{aligned}
    \phi^{r}_{l_{4}}(\theta\xrightarrow{}0^{+}) \sim& \left(\frac{r f^{\prime}-2f}{2r^{2}}\right)\left(\frac{1}{\theta}+\frac{1}{6}\theta+\mathcal{O}(\theta^{2})\right),\\
    \phi^{\theta}_{l_{4}}(\theta\xrightarrow{}0^{+}) \sim& \left(-\frac{\sqrt{f}}{r^{2}}\right)\left(\frac{1}{\theta^{2}}-\frac{1}{6}-\frac{7}{120}\theta^{2}+\mathcal{O}(\theta^{3})\right),
    \end{aligned}
\end{equation}
and
\begin{equation}
    \begin{aligned}
        \phi^{r}_{l_{2}}(\theta\xrightarrow{}\pi^{-}) \sim& \left(\frac{r f^{\prime}-2f}{2r^{2}}\right)\bigg(\frac{1}{\pi-\theta}+\frac{\pi-\theta}{6}\bigg.\\&\bigg.+\mathcal{O}\left((\pi-\theta)^{2}\right)\bigg),\\
    \phi^{\theta}_{l_{2}}(\theta\xrightarrow{}\pi^{-}) \sim& \left(\frac{\sqrt{f}}{r^{2}}\right)\bigg(\frac{1}{(\pi-\theta)^{2}}-\frac{1}{6}-\frac{7}{120}(\pi-\theta)^{2}\bigg.\\&\bigg.+\mathcal{O}\left((\pi-\theta)^{3}\right)\bigg).
    \end{aligned}
\end{equation}
Although $\phi^{r}_{l_{4}}, \phi^{\theta}_{l_{4}}, \phi^{r}_{l_{2}},$ and $\phi^{\theta}_{l_{2}}$ remain functions of $r$, this does not influence our asymptotic analysis. As $r$ increases from $r^{+}_{\text{h}}$, $r f^{\prime}-2f$ transitions from positive to negative values, causing $\phi^{r}$ to change sign from positive to negative. Despite this sign change in $\phi^{r}$, the behavior of $\phi^{\theta}$, as a higher-order term of $\phi^{r}$, dominates the direction of $\phi$. $\phi_{l_{4}}$ remains directed vertically downward, while $\phi_{l_{2}}$ remains directed vertically upward, i.e., $\Omega_{l_{4}}=-\pi/2$ and $\Omega_{l_{2}}=\pi/2$, leading to $\Delta\Omega_{l_{4}}=\Delta\Omega_{l_{2}}=0$.

In the analysis of three above cases, we have relied solely on the general properties of the metric functions for high-dimensional SSS asymptotically flat black holes, and thus the results naturally hold for such black holes.

When $r\xrightarrow{}+\infty$, we have
\begin{equation}
    \begin{aligned}
        \phi^{r}_{l_{1}}(r\xrightarrow{}+\infty) \sim& -\frac{1}{r^{2}\sin\theta}+\frac{D-1}{2}\frac{m}{r^{D-1}\sin\theta}\\&+\mathcal{O}\left(\frac{1}{r^{D}}\right),\\
        \phi^{\theta}_{l_{1}}(r\xrightarrow{}+\infty) \sim& -\frac{\cos\theta}{r^{2}\sin^{2}\theta}+\frac{m\cos\theta}{r^{D-1}\sin^{2}\theta}\\&+\mathcal{O}\left(\frac{1}{r^{D}}\right).
    \end{aligned}
\end{equation}
Thus, $\Omega_{l_1} = -\pi + \arctan(\cot \theta)$. As $\theta$ increases from 0 to $\pi$, $\Omega_{l_{1}}$ decreases smoothly from $-\pi/2$ to $-3\pi/2$, leading to $\Delta\Omega_{l_{1}}=-\pi$.

Combined with the straightforward angle analysis, we conclude that
$\Delta\Omega_{l_{1}\rightarrow l_{2}}=\Delta\Omega_{l_{4}\rightarrow l_{1}}=0$, $\Delta\Omega_{l_{2}\rightarrow l_{3}}=\Delta\Omega_{l_{3}\rightarrow l_{4}}=-\pi/2$, where $\Delta\Omega_{l_{a}\rightarrow l_{b}}$ represents the change in $\Omega$ at the junction of $l_{a}$ and $l_{b}$ along the positive direction.

Because $\Delta\Omega_{l_{2}}=\Delta\Omega_{l_{4}}=\Delta\Omega_{l_{1}\rightarrow l_{2}}=\Delta\Omega_{l_{4}\rightarrow l_{1}}=0$, we get $\Delta\Omega_{l_{1}\rightarrow l_{2}}+\Delta\Omega_{l_{2}}+\Delta\Omega_{l_{2}\rightarrow l_{3}}=\Delta\Omega_{l_{3}\rightarrow l_{4}}+\Delta\Omega_{l_{4}}+\Delta\Omega_{l_{4}\rightarrow l_{1}}=-\pi/2$. To demonstrate $\Delta\Omega_{l_{1}\rightarrow l_{2}}+\Delta\Omega_{l_{2}}+\Delta\Omega_{l_{2}\rightarrow l_{3}}$ and $\Delta\Omega_{l_{3}\rightarrow l_{4}}+\Delta\Omega_{l_{4}}+\Delta\Omega_{l_{4}\rightarrow l_{1}}$ are indeed equal to $-\pi/2$, rather than $-\pi/2+2k_{1}\pi$ $(k_{1}\in \mathbb{Z} \cap k_{1}\neq0)$, we construct a curve, $C_{1}$, as shown in Fig. \ref{Fig.1}. Due to the symmetry of $\phi$ with respect to $\theta=\pi/2$ and the positive direction, $\Delta\Omega_{l^{1}_{1}\rightarrow l^{1}_{2}}+\Delta\Omega_{l^{1}_{2}}+\Delta\Omega_{l^{1}_{2}\rightarrow l^{1}_{3}}$ equals $\Delta\Omega_{l^{1}_{3}\rightarrow l^{1}_{4}}+\Delta\Omega_{l^{1}_{4}}+\Delta\Omega_{l^{1}_{4}\rightarrow l^{1}_{1}}$. Thus, we focus on $\Delta\Omega_{l^{1}_{3}\rightarrow l^{1}_{4}}+\Delta\Omega_{l^{1}_{4}}+\Delta\Omega_{l^{1}_{4}\rightarrow l^{1}_{1}}$ for convenience. It is noted when $\theta_{0} \in (0, \pi/2)$, $\phi^{\theta_{0}}$ remains negative. Considering $\Omega_{l^{1}_{3}}(\theta_{0})=\Omega_{l_{3}}(\theta_{0})=0$ and $\Omega_{l^{1}_{1}}(\theta_{0})=\Omega_{l_{1}}(\theta_{0})=-\pi + \arctan(\cot \theta_{0})\in (-\pi/2, -\pi)$, we get $\Delta\Omega_{l^{1}_{3}\rightarrow l^{1}_{4}}+\Delta\Omega_{l^{1}_{4}}+\Delta\Omega_{l^{1}_{4}\rightarrow l^{1}_{1}} = -\pi + \arctan(\cot \theta_{0})$. As $\theta_{0}$ approaches $0$, $\Delta\Omega_{l^{1}_{3}\rightarrow l^{1}_{4}}+\Delta\Omega_{l^{1}_{4}}+\Delta\Omega_{l^{1}_{4}\rightarrow l^{1}_{1}}$ gradually approaches $-\pi/2$. This ultimately leads to $\Delta\Omega_{l_{3}\rightarrow l_{4}}+\Delta\Omega_{l_{4}}+\Delta\Omega_{l_{4}\rightarrow l_{1}}$ equals $-\pi/2$. Similarly, $\Delta\Omega_{l_{1}\rightarrow l_{2}}+\Delta\Omega_{l_{2}}+\Delta\Omega_{l_{2}\rightarrow l_{3}}$ equals $-\pi/2$.

In summary, the total topological charge
\begin{equation}
    \begin{aligned}
    Q=&\frac{1}{2\pi}(\Delta\Omega_{l_{1}}+\Delta\Omega_{l_{2}}+\Delta\Omega_{l_{3}}+\Delta\Omega_{l_{4}}+\Delta\Omega_{l_{1}\rightarrow l_{2}}\\&+\Delta\Omega_{l_{2}\rightarrow l_{3}}+\Delta\Omega_{l_{3}\rightarrow l_{4}}+\Delta\Omega_{l_{4}\rightarrow l_{1}})=-1.
    \end{aligned}
\end{equation}
It is evident that the total topological charge $Q$ is independent of the dimension $D$. This implies that the total topological charge of SSS asymptotically flat black holes in any dimension $( D \ge 5 )$ remains $-1$, indicating the existence of at least one standard PS outside the black hole horizon. In what follows, we proceed to test this conclusion with the regular black holes constructed from the pure gravity.

\section{Topological charge: high-dimensional regular black holes}
\label{tchdrbh}

Regular black holes are a class of black holes that undergo a specific construction to eliminate the central singularity associated with black holes. The quest to obtain regular black hole solutions is currently pursued through two main approaches. One method typically involves introducing exotic matter, which modifies the metric to a special form, leading to a regular black hole solution \cite{Dymnikova1,Ayon-Beato1,Ayon-Beato2,Bronnikov,Dymnikova2,Dymnikova3,Hayward}. The other approach involves modifying Einstein's theory of gravity, followed by solving the modified gravitational equations to obtain regular black hole solutions \cite{Oliva,Myers,Bambi,Frolov,Hennigar,Buoninfante,Simpson,Franzin}. Recently, Bueno, Cano, and Hennigar \cite{Bueno1} introduced an infinite series of higher-curvature corrections to Einstein's theory of gravity, successfully eliminating black hole singularities in any spacetime dimension $D \ge 5$. This theory achieves singularity elimination through purely gravitational mechanism, without the need for exotic matter fields. The resulting solutions belong to a class of quasi-topological gravity theories. By analyzing the asymptotic behavior at infinity, we find that these regular black holes are asymptotically flat black holes under certain conditions. We now aim to study the topology of the PSs for these black hole solutions.

We begin by introducing the purely gravitational approach used to construct regular black holes. The action of a quasi-topological gravity is given by
\begin{equation}
I_{QT}=\frac{1}{16 \pi G} \int \big(R+\sum_{n=2}^{n_{max}}\alpha_{n}Z_{n}\big)\sqrt{|g|} \,d^{D}x, \label{iqt}
\end{equation}
where $\alpha_{n}$ are arbitrary coupling constants with dimensions of length $2(n-1)$. $Z_{n}$ are the $n$-th-order quasi-topological densities \cite{Myers,Bueno2,Moreno}.

Considering a general SSS ansatz
\begin{equation}
ds^{2}=-N(r)^{2}f(r)dt^{2}+\frac{dr^{2}}{f(r)}+r^{2}d\varOmega^{2}_{D-2},
\end{equation}
the reduced action can be written as
\begin{equation}
I_{QT}[N,f]=\Omega_{D-2} \int\,dt \int\,dr L_{N,f},\label{iqtnf}
\end{equation}
where the effective Lagrangian $L_{N,f}$ is defined as
\begin{equation}
L_{N,f}=N(r)r^{D-2}\big(R+\sum_{n=2}^{n_{max}}\alpha_{n}Z_{n}\big)|_{_{N,f}}.
\end{equation}

Following \cite{Bueno3}, $L_{N,f}$ can be expanded as
\begin{equation}
L_{N,f}=NF^{\prime}_{0}+N^{\prime}F_{1}+N^{\prime\prime}F_{2}+\mathcal{O}\left(\frac{N^{\prime 2}}{N}\right),
\end{equation}
where $F^{\prime}_{0}=L_{N=1,f}$ , while $F_{1,2}$ are functions of $f(r)$ and its derivatives.

Performing an integration by parts, the action (\ref{iqtnf}) can be expressed as
\begin{equation}
    \begin{aligned}
    I_{QT}[N,f]=&\Omega_{D-2} \int\,dt \int\,dr \\& \left( N(F_{0}-F_{1}+F^{\prime}_{2})^{\prime}+\mathcal{O}\left(\frac{N^{\prime 2}}{N}\right)  \right).
    \end{aligned}
\end{equation}

The variations of $I_{QT}[N, f]$ with respect to $N$ and $f$ correspond respectively to the $tt$ and $rr$ components of the field equations, $\mathcal{E}_{ab}\equiv (1/\sqrt{|g|})(\delta I_{QT}/\delta g^{ab})$ \cite{Bueno4}: 
\begin{equation}
    \begin{aligned}
        \frac{1}{\Omega_{D-2}r^{D-2}}\frac{\delta I_{QT}[N,f]}{\delta N}=&\frac{2\mathcal{E}_{tt}}{fN^{2}},\label{ett}\\
        \frac{1}{\Omega_{D-2}r^{D-2}}\frac{\delta I_{QT}[N,f]}{\delta f}=&\frac{\mathcal{E}_{tt}}{Nf^{2}}+N\mathcal{E}_{rr},
    \end{aligned}
\end{equation}
one can obtain the equations of motion
\begin{eqnarray}
    \frac{dN}{dr}=0,\label{nnr}\\
    \frac{d}{dr}\left(F_{0}-F_{1}+F^{\prime}_{2}\right)=0.\label{eomf}
\end{eqnarray}

From (\ref{nnr}), it is clear that $N(r)$ is a quantity independent of $r$. Considering normalization of the time coordinate at infinity, it is reasonable to take $N(r)=1$. Under the condition $N(r)=1$, all quasi-topological densities can be written as
\begin{equation}
 \begin{aligned}
     r^{D-2}Z_{(p)}|_{N=1,f}=\frac{d}{dr}\big(&r^{D-1}\psi^{p-1}((2p-D)\psi\big.\\& \big.-2 p B) \big),\label{znn}
 \end{aligned}
\end{equation}
where $\psi=(1-f(r))/r^{2}$, $B=-f^{\prime}/(2r)$ and the set $\{Z_{(p)}; p=1,2,3,\cdots\}$ is obtained by normalizing $\{R, Z_{n}\}$ so that Eq. (\ref{znn}) is satisfied. The coefficients $\alpha_{n}$ change accordingly, but the notation is retained. The densities of different orders satisfying condition (\ref{znn}) are connected by a recurrence relation, ensuring that the condition remains valid for curvature densities of any order \cite{Bueno2}.

For quasi-topological gravity, the following condition is satisfied:
\begin{equation}
    \nabla^{a}P_{(p)abcd}|_{f}=0,
\end{equation}
where 
\begin{equation}    P_{(p)}^{abcd}\equiv\left(\frac{\partial Z_{(p)}}{\partial R_{abdc}}\right)_{g^{ef}}.
\end{equation}
The field equations reduce to
\begin{equation}
    {\mathcal{E}}_{a}^{b}|_{f}=\sum_{p=1}^{p_{max}}\alpha_p\left.\left(R_{ac}{}^{de}P_{(p)de}{}^{bc}-\frac{1}{2}\delta_{a}^{b}{Z}_{(p)}\right)\right|_{f}.\label{eab}
\end{equation}

Using Eqs. (\ref{ett}), (\ref{znn}), and (\ref{eab}), the equation of motion (\ref{eomf}) then becomes
\begin{equation}
    \begin{aligned}
    \frac{d}{dr}\left(F_{0}-F_{1}+F^{\prime}_{2}\right)=&\frac{d}{dr}\left( \sum_{p=1}^{p_{max}}\frac{D-2p}{2}\alpha_{p}r^{D-2p-1}(1-f)^{p} \right)\\=&\frac{d}{dr}\left( r^{D-1}h(\psi) \right)=0,
    \end{aligned}
\end{equation}
where $\alpha_{1}$ is equal to $1$
and
\begin{eqnarray}            h(\psi)=\psi+\sum_{n=2}^{n_{max}}\alpha_{n}\psi^{n}. \label{hpsi}
\end{eqnarray}
The factor $(D-2n)/2$ is absorbed into $\alpha_{n}$. In $F_{0}-F_{1}+F^{\prime}_{2}$, the terms $f^{\prime}$ and $f^{\prime\prime}$ cancel out, leaving only a purely algebraic equation that contains no $f^{\prime}$ and $f^{\prime\prime}$.

Considering $r^{D-1}h(\psi)$ is a quantity independent of $r$, we arrive
\begin{equation}
    h(\psi)=\frac{m}{r^{D-1}},
\end{equation}
where $m$ is an integration constant and is proportional to the Arnowitt-Deser-Misner (ADM) mass of the solution
\begin{equation}
    m=\frac{16\pi G}{(D-2)\Omega_{D-2}}M.
\end{equation}
If we consider $n_{max}$ is finite and $r \rightarrow 0$, $f(r)$ has the following asymptotic behaviour
{\begin{equation}
    f(r)=1-\left(\frac{m}{\alpha_{n_{max}}}\right)^{\frac{1}{n_{max}}}r^{2-\frac{(D-1)}{n_{max}}}+\cdots.
\end{equation}
Consider the following conditions \cite{Bueno1}
\begin{equation}
\alpha_{n} \ge 0 \ \forall \;n \ , \quad \displaystyle\lim_{n \rightarrow \infty}(\alpha_{n})^{\frac{1}{n}}= K > 0, \label{alphan}
\end{equation}
$h(\psi)$ is monotonic for $\psi > 0$ and has an inverse. To get regular black holes, $n_{max}$ should tend to infinity. Omitting the higher order terms, the asymptotic behaviour of $f(r)$ is given by
\begin{equation}
    f(r)=1-\frac{r^{2}}{K}.
\end{equation}
The solution has a regular core, which means the asymptotic behaviour of the metric function $f(r) =1-\mathcal{O}(r^{2})$ when $r \rightarrow 0$. The elimination of the singularity is evident from this asymptotic behaviour of $f(r)$.

The asymptotic behavior of these regular black holes at infinity is now considered. As $r \rightarrow \infty$, it is observed that $h(\psi)$, represented as $m/r^{D-1}$, tends towards zero. Through Eq. (\ref{hpsi}), it is deduced that $\psi$ along with $\sum_{n=2}^{n_{\text{max}}}\alpha_{n}\psi^{n}$ approaches zero. A condition weaker than that outlined in Eq. (\ref{alphan}) is utilized:
\begin{equation}
\alpha_{n} \ge 0 \ \forall n \ , \label{alphan1}
\end{equation}
ensuring $\psi = (1-f(r))/r^{2}$ tends towards zero. Consequently, the following asymptotic relationship is derived:
\begin{equation}
\psi + \sum_{n=2}^{n_{\text{max}}}\alpha_{n}\psi^{n} \sim \psi \sim \frac{1}{r^{D-1}}
\end{equation}
This indicates that $1-f(r) \sim 1/r^{D-3}$, resulting in $f(r)$ approaching 1. Thus, under the conditions stipulated in Eq. (\ref{alphan}), the regular black holes constructed through the pure gravity approach are characterized by asymptotic flatness.

Next, we investigate the topology of the PSs. Based on our theoretical derivation of the topology of the PSs for high-dimensional SSS asymptotically flat black holes in Sec. \ref{tchd}, it is reasonable to conjecture that the topological charge of the PSs for such regular black holes is -1. We now verify this conjecture using two characteristic examples.

\subsection{Hayward black hole: $\alpha_{n}=\alpha^{n-1}$}
\label{hbh}

Such parameter choice yields the Hayward black hole in high dimension
\begin{eqnarray}
    h(\psi)&=&\frac{\psi}{1-\alpha \psi},\\
    f(r)&=&1-\frac{mr^{2}}{r^{D-1}+\alpha m}.\label{fr1}
\end{eqnarray}

Substituting (\ref{fr1}) into (\ref{phirtheta}), we obtain
\begin{eqnarray}
    \phi^{r}&=&-\frac{1}{2(r^{D}+\alpha m r)^{2}}[2r^{2(D-1)}-(D-1)mr^{D+1}\\&&+4\alpha mr^{D-1}+2\alpha^{2} m^{2}]\csc \theta,\nonumber\\
    \phi^{\theta}&=&-\frac{\sqrt{r^{D-1}-mr^{2}+\alpha m}}{r^{D-3}+\alpha mr^{4}}\cot\theta \csc\theta.
\end{eqnarray}

\begin{figure}[htbp]
\center{\subfigure[]{\label{Fig.2a}
\includegraphics[width=4.1cm]{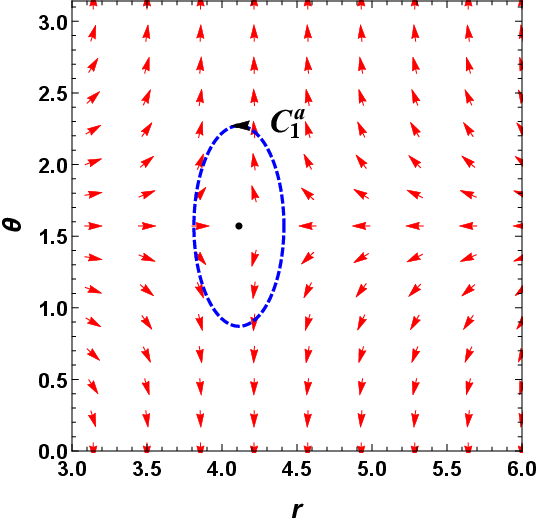}}
\subfigure[]{\label{Fig.2b}
\includegraphics[width=4.1cm]{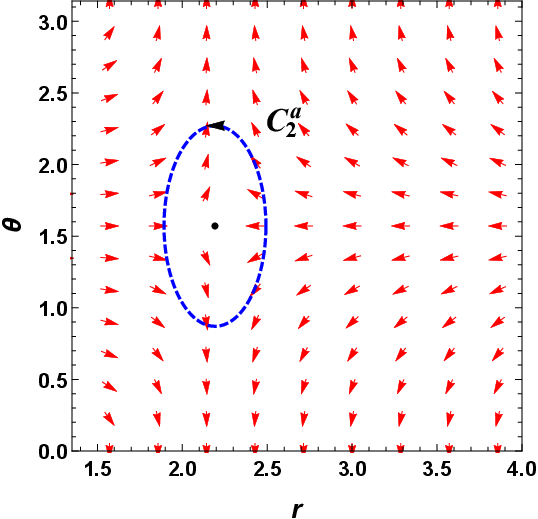}}
\subfigure[]{\label{Fig.2c}
\includegraphics[width=4.1cm]{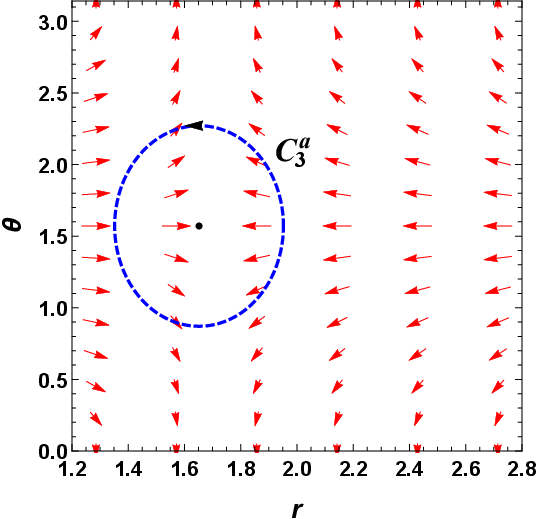}}
\subfigure[]{\label{Fig.2d}
\includegraphics[width=4.1cm]{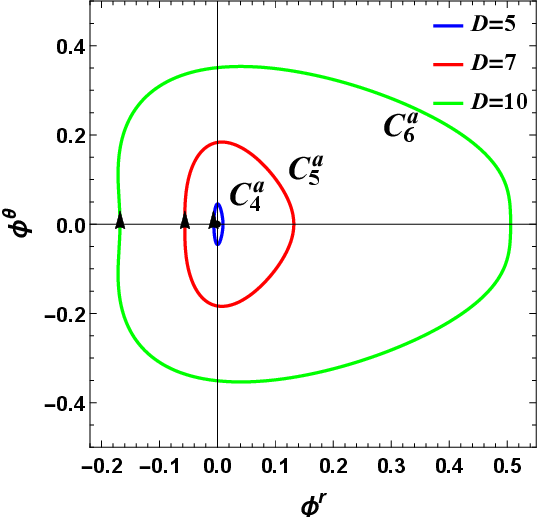}}}
\caption{The behaviour of the unit vector field $n$ in the $(r,\theta)$ plane in different dimensions, and the varying behaviour in $\phi$ space corresponds to the respective curves $C^{a}_{1}$, $C^{a}_{2}$ and $C^{a}_{3}$. The red arrow represents the direction of $n$, and the black dot represents the zero point of $n$. The blue dashed contour lines $C^{a}_{1}$, $C^{a}_{2}$ and $C^{a}_{3}$ are all closed ellipse centred on the zero point, and the solid contour lines $C^{a}_{4}$, $C^{a}_{5}$ and $C^{a}_{6}$ are the changes in the components $(\phi^{r} , \phi^{\theta})$ of $\phi$ along $C^{a}_{1}$, $C^{a}_{2}$ and $C^{a}_{3}$. (a) The unit vector field for the Hayward black hole with $D=5$, $\alpha = 1$ and $m=9$. (b) The unit vector field for the Hayward black hole with  $D=7$, $\alpha = 1$ and $m=9$. (c) The unit vector field for the Hayward black hole with $D=10$, $\alpha = 1$ and $m=9$. (d) The behaviour of $\phi$ in $\phi$ space for $C^{a}_{1}$, $C^{a}_{2}$ and $C^{a}_{3}$.}
  \label{Fig.2}
\end{figure}

Fixing the parameters $\alpha=1$ and $m=9$, we examine the topology of the PSs in different dimensions. The unit vectors $n$ with $D=5$, 7, and 10 are plotted in Figs. \ref{Fig.2a}, \ref{Fig.2b}, and \ref{Fig.2c}. Each of these images in Figs. \ref{Fig.2a}, \ref{Fig.2b}, and \ref{Fig.2c} contains only one zero point. By solving Eq. (\ref{rph}), each of the three cases yields only one solution, which coincides with the zero point of $n$, indicating that there is only one PS outside the outer horizon in these cases. A zero point corresponds to a PS. At this stage, the winding number equals the topological charge. As the dimension $D$ increases, the zero point shifts leftward, suggesting that the radius $r_{\text{ps}}$ of the PS decreases with $D$. Meanwhile, the radius $r_{\text{h}}$ of the black hole horizon decreases during this process. This leftward or rightward shift with increasing $D$ can be achieved by adjusting the parameters $\alpha$ and $m$. We describe the changes in the components $(\phi^{r}, \phi^{\theta})$ of $\phi$ in the vector space in Fig. \ref{Fig.2d}. The origin exactly corresponds to the zeros of $\phi$. When traversing $C^{a}_{1}$, $C^{a}_{2}$ and $C^{a}_{3}$ counterclockwise, the corresponding curves $C^{a}_{4}$, $C^{a}_{5}$ and $C^{a}_{6}$ are all clockwise, as shown by these black arrows in Figs. \ref{Fig.2a}, \ref{Fig.2b}, \ref{Fig.2c}, and \ref{Fig.2d}. The clockwise curve in the $\phi$ space is found to correspond to a negative topological charge,
\begin{equation}
    Q=-1,
\end{equation}
similar to the results in Ref. \cite{Wei5}.

\begin{figure}[htbp]
    \center{
    \includegraphics[width=6cm]{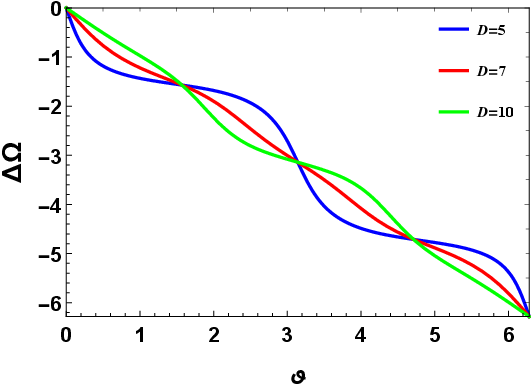}}
  \caption{$\Delta\Omega$ as a function of $\vartheta$ in different dimensions.}
  \label{Fig.3}
\end{figure}

We now calculate the topological charge numerically. Since the shape of the contour line that encloses the zero point does not affect the value of $Q$, we choose the elliptical contour lines for convenience.We can parameterize the contours $C^{a}_{1}$, $C^{a}_{2}$ and $C^{a}_{3}$ with the form (\ref{rthetavartheta}). When $D=5$, the zero point is located at $(4.11, \pi/2)$ and the parametric selection of $C^{a}_{1}$ is $(a, b ,r_{0})=(0.3, 0.7, 4.11)$. Similarly, the parameter selection of $C^{a}_{2}$ is $(0.3, 0.7, 2.19)$, for $C^{a}_{3}$ is $(0.3, 0.7, 1.65)$. Using (\ref{omegavartheta}), we calculate $\Delta \Omega$ along $C^{a}_{1}$, $C^{a}_{2}$ and $C^{a}_{3}$, as shown in Fig. \ref{Fig.3}. As $\vartheta$ increases from 0 to $2\pi$, all three curves monotonically decrease from 0 to $-2\pi$. Thus the topological charge $Q$ in all three cases is -1, indicating that they possess only one standard PS. It should be noted that if multiple PSs exist outside the horizon, $Q=-1$ represents at least one standard PS.

\subsection{Dymnikova-like black hole: $\alpha_{n}=\frac{\alpha^{n-1}}{n}$}

The metric function $f(r)$ of such a black hole is similar to the Dymnikova black hole \cite{Paul,Konoplya} and we call it the Dymnikova-like black hole. We have the metric functions
\begin{eqnarray}
    h(\psi)&=&-\frac{log(1-\alpha \psi)}{\alpha},\\
    f(r)&=&1-\frac{r^{2}}{\alpha}(1-e^{-\frac{\alpha m}{r^{D-1}}}).\label{fr2}
\end{eqnarray}

\begin{figure}[h]
\center{\subfigure[]{\label{Fig.4a}
\includegraphics[width=4.1cm]{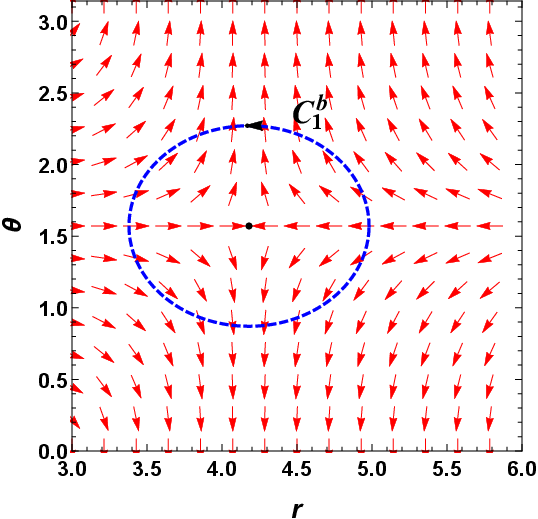}}
\subfigure[]{\label{Fig.4b}
\includegraphics[width=4.1cm]{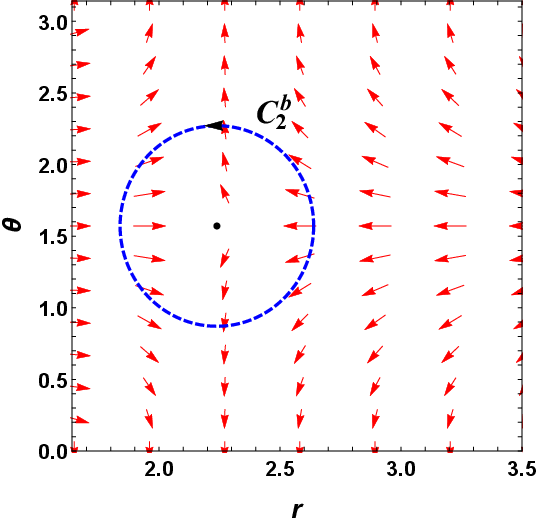}}
\subfigure[]{\label{Fig.4c}
\includegraphics[width=4.1cm]{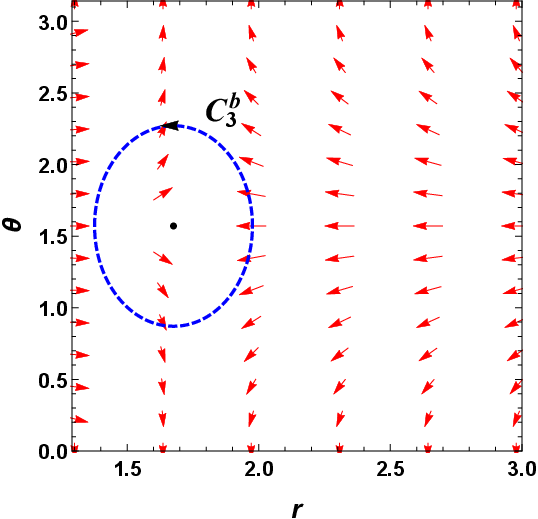}}
\subfigure[]{\label{Fig.4d}
\includegraphics[width=4.1cm]{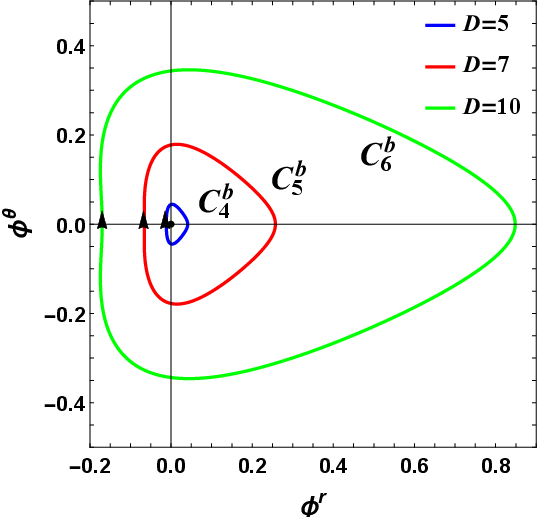}}}
\caption{The behaviour of the unit vector field $n$ in the $(r,\theta)$ plane in different dimensions, and the varying behaviour in $\phi$ space corresponds to the respective curves $C^{b}_{1}$, $C^{b}_{2}$ and $C^{b}_{3}$. The red arrow represents the direction of $n$, and the black dot represents the zero point of $n$. The blue dashed contour lines $C^{b}_{1}$, $C^{b}_{2}$ and $C^{b}_{3}$ are all closed ellipse centred on the zero point, and the solid contour lines $C^{b}_{4}$, $C^{b}_{5}$ and $C^{b}_{6}$ are the changes in the components $(\phi^{r} , \phi^{\theta})$ of $\phi$ along $C^{b}_{1}$, $C^{b}_{2}$ and $C^{b}_{3}$. (a) The unit vector field for the Dymnikova-like black hole with $D=5$, $\alpha = 1$ and $m=9$. (b) The unit vector field for the Dymnikova-like black hole with  $D=7$, $\alpha = 1$ and $m=9$. (c) The unit vector field for the Dymnikova-like black hole with $D=10$, $\alpha = 1$ and $m=9$. (d) The behaviour of $\phi$ in $\phi$ space for $C^{b}_{1}$, $C^{b}_{2}$ and $C^{b}_{3}$.}
  \label{Fig.4}
\end{figure}

Substituting Eq. (\ref{fr2}) into Eq. (\ref{phirtheta}), we obtain
\begin{equation}
\begin{aligned}
\phi^{r}&=-\frac{2r^{D-3}-(D-1)me^{-\frac{\alpha m}{r^{D-1}}}}{2r^{D-1}}\csc\theta,\\
\phi^{\theta}&=-\frac{\sqrt{(e^{-\frac{\alpha m}{r^{D-1}} }-1)r^{2}+\alpha}}{\sqrt{\alpha}r^{2}}\cot\theta \csc\theta.
\end{aligned}
\end{equation}

\begin{figure}[htbp]
    \center{
    \includegraphics[width=6cm]{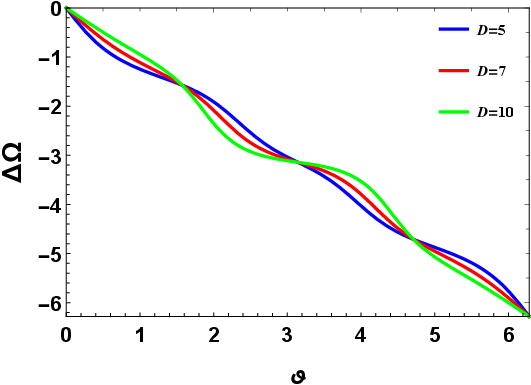}}
 \caption{$\Delta\Omega$ as a function of $\vartheta$ in different dimensions.}
  \label{Fig.5}
\end{figure}

Fixing the parameters $\alpha=1$ and $m=9$, we examine the topology of the PSs in different dimensions. The unit vectors $n$ with $D=5$, 7, and 10 are plotted in Figs. \ref{Fig.4a}, \ref{Fig.4b}, and \ref{Fig.4c}, respectively. These all have only one zero point in Figs. \ref{Fig.4a}, \ref{Fig.4b}, and \ref{Fig.4c}. By solving Eq. (\ref{rph}), each of the three cases yields only one solution, which coincides with the zero point of $n$. It indicates there is only one PS outside the outer horizon in these cases. The winding number equals the topological charge. The phenomena observed when fixing certain parameters $\alpha$ and $m$ are similar to that of the Hayward black hole. The zero position shifts to the left as the dimension $D$ increases. When traversing $C^{b}_{1}$, $C^{b}_{2}$ and $C^{b}_{3}$ counterclockwise, the corresponding curves $C^{b}_{4}$, $C^{b}_{5}$ and $C^{b}_{6}$ are all clockwise, as shown by these black arrows in Figs. \ref{Fig.4a}, \ref{Fig.4b}, \ref{Fig.4c}, and \ref{Fig.4d}. The clockwise curves $C^{b}_{4}$, $C^{b}_{5}$ and $C^{b}_{6}$ indicate
\begin{equation}
    Q=-1.
\end{equation}

Now we calculate the topological charge numerically. Using Eq. (\ref{rthetavartheta}), we can parameterize $C^{b}_{1}$, $C^{b}_{2}$ and $C^{b}_{3}$. The parameter selections are as follows: ($a$, $b$, $r_0$)=$(0.8, 0.7, 4.18)$, $(0.4, 0.7, 2.24)$, and $(0.3, 0.7, 1.68)$ for $C^{b}_{1}$, $C^{b}_{2}$, and $C^{b}_{3}$. Using (\ref{omegavartheta}), we calculate $\Delta \Omega$ along $C^{b}_{1}$, $C^{b}_{2}$ and $C^{b}_{3}$ shown in Fig. \ref{Fig.5}. As $\vartheta$ increases from 0 to $2\pi$, all these three curves monotonically decrease from 0 to $-2\pi$. Thus the topological charge $Q$ in all three cases is -1. This indicates that they possess only one standard PS.

The two examples with different choices of $\alpha_{n}$ serve as a validation of the previous conclusion that the topological charge of the PSs of a SSS asymptotically flat black hole is always -1, independent of the dimension ($D \ge 5$).

It is also noteworthy that this purely gravitational approach constructs not only regular black holes but also black holes with singularity under the condition (\ref{alphan1}). Regardless of whether $n$ tends to infinity, under the condition (\ref{alphan1}), these constructed black holes are asymptotically flat. Our conclusion applies to all of them. The topological charge remains -1, independent of whether a singularity exists inside the black hole.

\section{Discussions and conclusion}
\label{Conclusion}

In this paper, we studied the topology of the PSs of higher-dimensional SSS asymptotically flat black holes $(D \ge 5)$. According to Duan's $\phi$-mapping topological current theory, we determine the topological charge for the high-dimensional SSS asymptotically flat black holes by analyzing the asymptotic behavior. The topological charge for such black holes remains -1, indicating that these black holes possess at least one standard PS, and thus they belong to the same $Q=-1$ topological class.

Regular black holes constructed using the purely gravitational method are investigated. Condition (\ref{alphan1}) ensures the asymptotic flatness of these regular black holes, making it logical that we make the assumption that the topological charge $Q=-1$ for these regular black holes. We provided two examples of such regular black holes: Hayward black holes and Dymnikova-like black holes. In all cases, we fixed the parameters $m$ and $\alpha$ and examined their behavior across dimensional variations. The unit vector field $n$ in the $(r, \theta)$ plane has a single zero point outside the outer horizon in all cases. This zero point shifts leftward with increasing dimension, corresponding to a decrease in the radius of the PS. However, this shift is not essential. By adjusting the parameters $\alpha$ and $m$, we can control the direction of the shift. The closed curves in the $(r, \theta)$ plane containing the zeros map to closed curves in the $\phi$ space that contain the origin. We found that traversing the former counterclockwise is equivalent to traversing a clockwise circle in $\phi$ space. Referring to the definitions about $\beta_{m}$ and $\eta_{m}$ in Sec. \ref{tc}, we observed that this demonstrates the topological charge is -1. Numerical calculations futher confirm that the topological charge is indeed -1. These regular black holes belong to the same $Q=-1$ topological class.

Our study confirms that the topological charge for high-dimensional SSS asymptotically flat black holes always equals -1, indicating that at least one standard PS exists outside them. This topological approach offers an efficient method for analyzing the PSs of black holes. We expect that it will be applied to investigate other high-dimensional black holes with varying asymptotic behaviors. Such approach will serve as a valuable guide for studying the PSs of high-dimensional black holes.

\section*{Acknowledgements}
This work was supported by the National Natural Science Foundation of China (Grants No. 12475055, and No. 12247101), the Fundamental Research Funds for the Central Universities (Grant No. lzujbky-2025-jdzx07), and the Natural Science Foundation of Gansu Province (No. 22JR5RA389, No.25JRRA799).

\end{document}